\input harvmac
\overfullrule=0pt
\def\Title#1#2{\rightline{#1}\ifx\answ\bigans\nopagenumbers\pageno0\vskip1in
\else\pageno1\vskip.8in\fi \centerline{\titlefont #2}\vskip .5in}

\lref\peet{
  A.~W.~Peet,
  ``The Bekenstein formula and string theory (N-brane theory),''
  Class.\ Quant.\ Grav.\  {\bf 15}, 3291 (1998)
  [arXiv:hep-th/9712253].
}
\lref\haw{
  S.~W.~Hawking,
  ``Particle Creation By Black Holes,''
  Commun.\ Math.\ Phys.\  {\bf 43}, 199 (1975)
  [Erratum-ibid.\  {\bf 46}, 206 (1976)].
}

\lref\bek{
  J.~D.~Bekenstein,
  ``Black holes and entropy,''
  Phys.\ Rev.\  D {\bf 7}, 2333 (1973).
}
\lref\lt{D.~A.~Lowe and L.~Thorlacius,
  ``AdS/CFT and the information paradox,''
  Phys.\ Rev.\  D {\bf 60}, 104012 (1999)
  [arXiv:hep-th/9903237].}
\lref\clp{S.\ Carlip, Phys. Rev. {\bf D51} (1995) 632.}
\lref\ascv{A.\ Strominger and C.\ Vafa,
``Microscopic Origin of the Bekenstein-Hawking Entropy"
 { Phys. Lett.} {\bf B379}
(1996) 99, hep-th/9601029.}
\lref\nhbh{
  A.~Strominger,
  ``Black hole entropy from near-horizon microstates,''
  JHEP {\bf 9802}, 009 (1998)
  [arXiv:hep-th/9712251].}
 \lref\rov {
  C.~Rovelli,
  ``Black hole entropy from loop quantum gravity,''
  Phys.\ Rev.\ Lett.\  {\bf 77}, 3288 (1996)
  [arXiv:gr-qc/9603063].
}
\lref\jm{
  J.~M.~Maldacena,
  ``The large N limit of superconformal field theories and supergravity,''
  Adv.\ Theor.\ Math.\ Phys.\  {\bf 2}, 231 (1998)
  [Int.\ J.\ Theor.\ Phys.\  {\bf 38}, 1113 (1999)]
  [arXiv:hep-th/9711200].}
\lref\btz{M.\ Banados, C.\ Teitelboim and J.\ Zanelli, {\sl
Phys. Rev. Lett.} {\bf 69} (1992) 1849.}\lref\jc{J. A. Cardy, Nucl. Phys. {\bf B270} (1986) 186.}
\lref\jbmh{J.\ D.\ Brown and M.\
Henneaux, { Comm. Math. Phys.} {\bf 104} (1986) 207.}
\lref\jmas{J. Maldacena and A. Strominger, Phys.Rev. {\bf D55} (1997) 861.} 
\lref\sussum{
  L.~Susskind and J.~Uglum,
  ``Black hole entropy in canonical quantum gravity and superstring theory,''
  Phys.\ Rev.\  D {\bf 50}, 2700 (1994)
  [arXiv:hep-th/9401070].}
\lref\jacb{T.~Jacobson,
  ``Black hole entropy and induced gravity,''
  arXiv:gr-qc/9404039,   D.~N.~Kabat,
  ``Black hole entropy and entropy of entanglement,''
  Nucl.\ Phys.\  B {\bf 453}, 281 (1995)
  [arXiv:hep-th/9503016], V.~P.~Frolov and D.~V.~Fursaev,
  ``Thermal fields, entropy, and black holes,''
  Class.\ Quant.\ Grav.\  {\bf 15}, 2041 (1998)
  [arXiv:hep-th/9802010], 
T.~Jacobson and R.~Parentani, 
  ``Black hole entanglement entropy regularized in a freely falling frame,''
  Phys.\ Rev.\  D {\bf 76}, 024006 (2007)
  [arXiv:hep-th/0703233].}
 \lref\all{ T.~M.~Fiola, J.~Preskill, A.~Strominger and S.~P.~Trivedi,
  ``Black hole thermodynamics and information loss in two-dimensions,''
  Phys.\ Rev.\  D {\bf 50}, 3987 (1994)
  [arXiv:hep-th/9403137].}
\lref\sak{
  A.~D.~Sakharov,
  ``Vacuum quantum fluctuations in curved space and the theory of
  gravitation,''
  Sov.\ Phys.\ Dokl.\  {\bf 12}, 1040 (1968)
  [Dokl.\ Akad.\ Nauk Ser.\ Fiz.\  {\bf 177}, 70 (1967\ SOPUA,34,394.1991\ GRGVA,32,365-367.2000)].}
\lref\carlip{S.~Carlip,
  ``Near-horizon conformal symmetry and black hole entropy,''
  Phys.\ Rev.\ Lett.\  {\bf 88}, 241301 (2002)
  [arXiv:gr-qc/0203001];
  ``Black Hole Entropy and the Problem of Universality,''
  arXiv:0807.4192 [gr-qc].}
\lref\mhs{
  S.~Hawking, J.~M.~Maldacena and A.~Strominger,
  ``DeSitter entropy, quantum entanglement and AdS/CFT,''
  JHEP {\bf 0105}, 001 (2001)
  [arXiv:hep-th/0002145], R.~Emparan,
  ``Black hole entropy as entanglement entropy: A holographic derivation,''
  JHEP {\bf 0606}, 012 (2006),
  [arXiv:hep-th/0603081],  M.~Cadoni,
  ``Induced gravity and entanglement entropy of 2D black holes,''
  PoS {\bf QG-PH}, 013 (2007)
  [arXiv:0709.0163 [hep-th]].}
\lref\tak{
  T.~Nishioka, S.~Ryu and T.~Takayanagi,
  ``Holographic Entanglement Entropy: An Overview,''
  arXiv:0905.0932 [hep-th].}
\lref\sred{
  L.~Bombelli, R.~K.~Koul, J.~H.~Lee and R.~D.~Sorkin,
  ``A Quantum Source of Entropy for Black Holes,''
  Phys.\ Rev.\  D {\bf 34}, 373 (1986), R.~D.~Sorkin,
  ``Toward A Proof Of Entropy Increase In The Presence Of Quantum Black
  Holes,''
  Phys.\ Rev.\ Lett.\  {\bf 56}, 1885 (1986), 
  M.~Srednicki,
  ``Entropy and area,''
  Phys.\ Rev.\ Lett.\  {\bf 71}, 666 (1993)
  [arXiv:hep-th/9303048].}
  \lref\eva{
  E.~Silverstein,
  ``Simple de Sitter Solutions,''
  Phys.\ Rev.\  D {\bf 77}, 106006 (2008)
  [arXiv:0712.1196 [hep-th]].}
\lref\bms{
  R.~Bousso, A.~Maloney and A.~Strominger,
  ``Conformal vacua and entropy in de Sitter space,''
  Phys.\ Rev.\  D {\bf 65}, 104039 (2002)
  [arXiv:hep-th/0112218].}
\lref\asds{
  E.~Witten,
  ``Quantum gravity in de Sitter space,''
  arXiv:hep-th/0106109, 
  A.~Strominger,
  ``The dS/CFT correspondence,''
  JHEP {\bf 0110}, 034 (2001)
  [arXiv:hep-th/0106113];  ``Inflation and the dS/CFT correspondence,''
  JHEP {\bf 0111}, 049 (2001)
  [arXiv:hep-th/0110087],
  A.~Guijosa and D.~A.~Lowe,
  ``A new twist on dS/CFT,''
  Phys.\ Rev.\  D {\bf 69}, 106008 (2004)
  [arXiv:hep-th/0312282].
}
\lref\witcsg{ A.~Achucarro and P.~K.~Townsend,
  ``A Chern-Simons Action for Three-Dimensional anti-De Sitter Supergravity
  Theories,''
  Phys.\ Lett.\  B {\bf 180}, 89 (1986); 
  E.~Witten,
  ``(2+1)-Dimensional Gravity as an Exactly Soluble System,''
  Nucl.\ Phys.\  B {\bf 311}, 46 (1988).}
\lref\fjrl{
  J.~Fjelstad and S.~Hwang,
  ``Sectors of solutions in three-dimensional gravity and black holes,''
  Nucl.\ Phys.\  B {\bf 628}, 331 (2002)
  [arXiv:hep-th/0110235].}
\lref\witrec{
  E.~Witten,
  ``Three-Dimensional Gravity Revisited,''
  arXiv:0706.3359 [hep-th].}
\lref\witmal{
  A.~Maloney and E.~Witten,
  ``Quantum Gravity Partition Functions in Three Dimensions,''
  arXiv:0712.0155 [hep-th], 
  S.~Giombi, A.~Maloney and X.~Yin,
  ``One-loop Partition Functions of 3D Gravity,''
  JHEP {\bf 0808}, 007 (2008)
  [arXiv:0804.1773 [hep-th]].}
\lref\lss{
  W.~Li, W.~Song and A.~Strominger,
  ``Chiral Gravity in Three Dimensions,''
  JHEP {\bf 0804}, 082 (2008)
  [arXiv:0801.4566 [hep-th]].}
\lref\mss{
  A.~Maloney, W.~Song and A.~Strominger,
  ``Chiral Gravity, Log Gravity and Extremal CFT,''
  arXiv:0903.4573 [hep-th].
}
\lref\fks{
  S.~Ferrara, R.~Kallosh and A.~Strominger,
  ``N=2 extremal black holes,''
  Phys.\ Rev.\  D {\bf 52}, 5412 (1995)
  [arXiv:hep-th/9508072].}
\lref\sen{A.~Dabholkar, A.~Sen and S.~P.~Trivedi,
  ``Black hole microstates and attractor without supersymmetry,''
  JHEP {\bf 0701}, 096 (2007)
  [arXiv:hep-th/0611143].}
\lref\ghss{
  M.~Guica, T.~Hartman, W.~Song and A.~Strominger,
  ``The Kerr/CFT Correspondence,''
  arXiv:0809.4266 [hep-th].}
\lref\hinf{S.~W.~Hawking,
  ``Breakdown Of Predictability In Gravitational Collapse,''
  Phys.\ Rev.\  D {\bf 14}, 2460 (1976).}
\lref\page{
  D.~N.~Page,
  ``Is Black Hole Evaporation Predictable?,''
  Phys.\ Rev.\ Lett.\  {\bf 44}, 301 (1980).}
\lref\jmc{J.~M.~Maldacena,
  ``Eternal black holes in Anti-de-Sitter,''
  JHEP {\bf 0304}, 021 (2003)
  [arXiv:hep-th/0106112].}
\lref\spike{   L.~Fidkowski, V.~Hubeny, M.~Kleban and S.~Shenker,
  ``The black hole singularity in AdS/CFT,''
  JHEP {\bf 0402}, 014 (2004)
  [arXiv:hep-th/0306170]; G.~Festuccia and H.~Liu,
  ``Excursions beyond the horizon: Black hole singularities in Yang-Mills
  theories. I,''
  JHEP {\bf 0604}, 044 (2006)
  [arXiv:hep-th/0506202];  N.~Iizuka, T.~Okuda and J.~Polchinski,
  ``Matrix Models for the Black Hole Information Paradox,''
  arXiv:0808.0530 [hep-th].}
\lref\hnew{
  S.~W.~Hawking,
  ``Information Loss in Black Holes,''
  Phys.\ Rev.\  D {\bf 72}, 084013 (2005)
  [arXiv:hep-th/0507171].}

\Title{\vbox{\baselineskip12pt}}
{\vbox{\centerline {Five Problems in Quantum Gravity}}}
\centerline{Andrew Strominger\footnote{*}{strominger@physics.harvard.edu}}

\bigskip\centerline{Center for the Fundamental Laws of Nature}
\centerline{Harvard University}\centerline{Cambridge, MA USA}
%
%

\def\[{\left [}
\def\]{\right ]}
\def\({\left (}
\def\){\right )}

\vskip .3in

\centerline{\bf Abstract}
We present five open problems in quantum gravity which one might reasonably hope to solve in the next decade. Hints appearing in the literature are summarized for each one.

\smallskip
\noindent
\Date{{\it Based on lectures given at the 2008 Cargese
Summer School.} }
\listtoc
\writetoc
\newsec{Introduction}
Reconciling quantum mechanics and general relativity is one of the great scientific challenges of our time. A definitive resolution will undoubtedly require both experimental and theoretical advances. At present, hopes for relevant physical experiments are distant. On the other hand, we are rich in puzzles and gedanken experiments which can help us understand the
severe constraints imposed by theoretical consistency. The results of these gedanken experiments have led to dramatic theoretical advances over the last several decades, with a notable role played by string theory.  Along the way, the problem of quantum gravity has revealed an unanticipated  depth  and richness, reaching into, tying together and sometimes solving problems in disparate areas of math and physics. Compelling new paradigms have been suggested for the structure of the universe around us.  At the same time, it seems clear that what we have learned so far is only the tip of the iceberg and there is much more to come.

In this lecture, I will describe five problems of varying difficulty in quantum gravity that strike me as  ripe for attack and might conceivably be solved in the next decade.\foot{Each of the five problems discussed here has hundreds or thousands of relevant references, so it is impossible to give a comprehensive list. Those given are simply meant to be representative and provide the reader with an entry to the literature.
}  The problems here are ``sharp" in the sense that there is a definite number or function involved.\foot{Should the reader wish to suggest additional short hints or sharp problems, a communication to the author for use in a revised or expanded version of this note would be appreciated.} A successful computation of the number/function from some theoretical starting point is then a good indicator we are on the right track.  Perhaps consideration of these problems will help us find the rest of  the iceberg.

\newsec{Universality of the Bekenstein-Hawking area-entropy law}
 \subsec{The problem}    The Bekenstein-Hawking area law \refs{\bek,\haw}
\eqn\dft{S_{BH}= {{ \rm Area} \over 4  \hbar G}}
applies universally to all horizons - cosmic, black hole and observer -in general relativity. Consistency with the generalized laws of thermodynamics requires that this entropy can be accounted for by counting some kind of quantum microstates.  This accounting  was achieved,
including the ${1\over 4}$ prefactor,  
for certain five-dimensional supersymmetric black holes in string theory \ascv, and then generalized to a wide variety of contexts, including many which do not involve string theory \peet. However in this construction and its generalizations, the fact that the entropy is proportional to the area comes out only at the last step of a long computation. It is not obvious why this should always turn out to be so. A simple universal relation like \dft\  demands a simple universal explanation.  The problem is to find it. 
\subsec{Some hints}
 
\noindent {\bf (i)} If we tile the horizon with Planck-sized cells, and assign one degree of freedom to each cell, then the entropy, which is extensive, will go like the area. This suggests that the microstates can be described as living  on the horizon itself.  The hard part is to naturally get the ${1 \over 4}$ from such a picture \rov.

\noindent{\bf (ii)} The entropy has a one-loop correction, proportional to $\hbar^0$, which is finite when expressed in terms of the one-loop corrected 
Newton's constant \sussum, modulo some subtleties \jacb. This correction can be viewed as the entanglement entropy of the quantum states of the quantum fields inside and outside the horizon. It is dominated by short-wavelength modes and is therefore naturally 
proportional to the area \sred.  If gravity is induced \sak, which means that Newton's constant is zero at tree level and arises as a one loop correction, then the entanglement entropy  is responsible for $all$ of the entropy, and reproduces the area law with the correct coefficient \refs{\jacb,\all}.  This might in fact be the case in string theory, where the Einstein action is induced at one loop from open strings, but this notion has yet to be made precise.  Recent progress \tak\ has revealed a rich 
holographic relation between entanglement entropy and minimal surfaces including horizons. Related observations appear in \mhs.

\noindent{\bf(iii)} A universal relation in general relativity could be the image of a universal relation in statistical mechanics. One such candidate is the  Cardy formula \jc\ which relates the asymptotic growth of states of a 2D CFT to its central charge. This formula was the basis of the original stringy computation in \ascv.  It was later shown \nhbh\ that the 
Cardy/Area relation transcends its stringy origins, and follows in many cases simply from properties of the diffeomorphism group \jbmh.  Could the area law $always$ be the Cardy formula? One suggestion along these lines \carlip\ is that there 
might be some kind of universal 2D CFT, where the conformal transformations act in the $(r,t)$ plane, for all horizons. 
Another hint is that the general formula for 4D Kerr-Newman entropy can be written in the Cardy form
$S_{BH}=2\pi(\sqrt{c_L\bar L_0/6}+\sqrt{c_R L_0/6})$ for $c_L=c_R=6,~~~L_0=(M^4-Q^2M^2-J^2), ~~~\bar L_0=
(M^2-\half Q^2)^2$.

\newsec{de Sitter entropy}
 \subsec{The problem}  de Sitter space has an event horizon with thermodynamic properties described by the area law \dft.  The problem is to reproduce the de Sitter entropy by microstate counting. The location of the horizon is observer dependent, like the horizon in Rindler space which also obeys an area law. Since the de Sitter area law is numerically the same as the black hole area law, they must have a common explanation. It is very hard to imagine how this can work. For the black hole, we can at least say approximately ``where" the microstates are located; there is an object with states to be counted.  The object whose states we are supposed to count in de Sitter space is more elusive. 
 \subsec{Some hints}
 \noindent{\bf(i)} One way to proceed is to try to find de Sitter space as a solution of string theory and then, 
 as in the black hole problem \ascv, find a duality transformation which maps it to a quantum system whose microstates can be counted. de Sitter solutions are difficult, but not impossible, to describe because they are never supersymmetric. Some recent rather simple constructions can be found in \eva. 
 
 \noindent{\bf(ii)} de Sitter space has an asymptotic boundary  which is similar in many respects to that of anti-de Sitter space, but differs in that the boundaries are at timelike rather than spacelike infinity. The similarity  suggests the possibility of
a holographic dS/CFT correspondence \asds\ along the lines of the AdS/CFT correspondence \refs{\jm,\jbmh}. 
The microstates might then be counted in the dual CFT. Some tantalizing numerological evidence for this was found in 
\bms.

\newsec{Partition function of 3D AdS-Einstein gravity}
\subsec{The problem}  In three dimensions, all solutions of Einstein gravity with a negative cosmological constant are locally, but not necessarily globally, AdS$_3$. There are therefore no local degrees of freedom, and one might think the theory is too trivial to be interesting. On the other hand, it contains black hole solutions \btz, so the quantum version, if it indeed exists,  must at least be rich enough to account for the black hole microstates.  A sharp question is to compute the quantum partition function of pure 3D AdS-Einstein gravity as a function of Newton's constant $G$ and the cosmological constant $-\ell^{-2}$. It is surprising that pure 3D gravity has been studied for  decades by now and it is still not known if there is a consistent quantum version. 
 \subsec{Some hints}  
   \noindent{\bf(i)} In \witcsg, it was proposed, based on local equivalence of the equations of motion, that  pure AdS$_3$ gravity can be solved by rewriting it as  a Chern-Simons gauge theory and then using the holographic duality to a (reduced) boundary WZW model. However this proposal has run into problems \witrec .
 One way of stating the problem is that the euclidean partition function $Z$ constructed this way fails to be invariant under modular transformations. Since these are large diffeomorphisms this is a quantum  anomaly. A closely related statement is that the WZW microstates are not numerous enough to account for the black hole entropy (although see \fjrl). So it seems that in fact the Chern-Simons gauge theory is not quantum equivalent to pure gravity - perhaps because it includes singular field configurations with vanishing metric. Nevertheless, the connection  to Chern-Simons gauge theory smells like an important hint.

\noindent{\bf(ii)} On very general grounds \jbmh, we expect that 3D AdS gravity should be dual to a 2D CFT with central charge 
$c={3\ell \over 2 G}$. Solving the theory is equivalent to specifying this CFT. It was suggested in \witrec\ that, rather than 
directly quantizing the Einstein-Hilbert action,  this CFT might simply be deduced by various consistency requirements. Namely, the central charge must be $c={3\ell \over 2 G}$, $Z$ must be modular invariant (since these are large diffeomorphisms) and its pole structure must reflect the fact that there are no perturbative excitations. Adding the additional $assumption$ of holomorphic factorization (i.e. decoupling of the left and right movers in the CFT), it was shown \witrec\ that $Z$ is uniquely determined to be a certain modular form $Z_{ext}$. Unfortunately $Z_{ext}$ does not agree with the Euclidean sum-over-geometries \witmal\  which indicates that the assumption is not valid for pure gravity.\foot{ Interestingly, the assumption becomes a consistency requirement for chiral gravity \lss\ whose action contains an additional gravitational Chern-Simons term. Therefore if quantum chiral gravity exists, the argument of \witrec\ can be applied and its partition function must be $Z_{ext}$. This conclusion does agree with the sum-over-geometries \mss.}  Modular invariance and the restriction on the pole structure are still strong, if not uniquely determining, hints on the form of $Z$ for pure gravity. Determining $Z$ for pure 3D quantum Einstein gravity - if it exists - is an important open problem. 

\newsec{Extreme Kerr-Newman}
\subsec{The problem}
The easiest kind of black holes to understand are the stable, charged, supersymmetric ones.  Perhaps the hardest are the neutral Schwarzchild black holes. There are no useful parameters to expand in, and they represent an excited quantum system which is decaying.  An intermediate step between the easy supersymmetric and the difficult Schwarzchild black holes is the two parameter family of 
extreme Kerr-Newman black holes. These have zero Hawking temperature as well as variable parameters and so are perhaps simpler to understand than Schwarzchild. The problem is to give a statistical accounting of the entropy $S_{EKN}=\pi\sqrt{Q^4+4J^2}$ of an extreme charge $Q$ spin $J$ Kerr-Newman black hole. 
\subsec{Some hints}
\noindent{\bf(i)} It has been understood for some time that the near horizon region of supersymmetric black holes
is governed by an attractor mechanism, in which the geometry is determined by the charges independently of the asymptotic data of the spacetime \fks. This mechanism is essential in order for the area-law entropy to be an intrinsic property of the black hole. Recently it has been understood that the attractor mechanism operates for 
$all$ extreme black holes supersymmetric or not \sen, implying that the number of quantum microstates is invariant under at least some changes in the external parameters. 

\noindent{\bf(ii)} Interestingly, the attractor geometry always contains an enhanced $SL(2,R)$ isometry, which becomes a (warped) AdS$_3$ when the fibration of the $U(1)$ of angular momentum or  electric charge is included.  This suggests that extreme black holes may  always be  related to 2D CFTs. Indeed, since these lectures were given, progress has been made along these lines \ghss. 
\newsec{Information release}
\subsec{The problem}In the seventies \hinf\ Hawking gave a very simple and beautiful argument that 
information is destroyed by black holes. As our understanding has progressed, despite the simplicity of the argument, Hawking's original scenario has seemed less and less plausible \hnew. A variety of papers have appeared pointing to possible loopholes, but there does not appear to be a general consensus on how, where and why Hawking's  argument fails.  The problem is to explain how the information is released from a decaying black hole.

A sharp question, which may serve to clarify the discussion, is:  What is the rate of information return  to ${\cal I}^+$ of an evaporating black hole? To be more precise, a mixed state $\rho(t^-)$ on ${\cal I}^+$ at retarded time $t^-$ can be defined by tracing over 
the portion of the full state on ${\cal I}^+$ with support after $t^-$.  One can then compute the entanglement entropy $S_{ent}(t^-)$ of $\rho$ from the usual expression  $-\tr \rho \ln \rho$.  If the information is all returned, then $S_{ent}(\pm \infty)=0$. The problem is to compute the function $S_{ent}(t^-)$. If we truly understand black hole dynamics, we should be able to compute this function. 

While there are many possibilities, several candidates stand out:

\noindent{\bf(1) Bad question:} The question/answer is for some reason ill-defined.

\noindent{\bf(2) Information destruction:} $ S_{ent}(t^-)$ grows monotonically as expected from Hawking's analysis for a time of order $M^3$ (for 4D Schwarzchild), and then stays constant when  the black hole evaporates to zero radius and disappears.  Information is destroyed. 

\noindent{\bf(3) Long-lived remnant:} $S_{ent}(t^-)$ grows monotonically as expected from Hawking's analysis for a time of order $M^3$, and then slowly decreases back to zero over a time of order $M^4$ until the black hole becomes Planckian.\foot{In a time $M^4$ there will be an energy of order one in radially propagating infrared modes spread over a region of size $M^4$. This is the size needed to accommodate an entropy of order $M^2$. } The information is slowly released in infrared quanta with energy of order $M^{-4}$  by a conventional long-lived remnant. 

\noindent{\bf(4) Non-local remnant:} $S_{ent}(t^-)$ grows monotonically as expected from Hawking's analysis for a time of order $M^3$, and then decreases back to zero over a time of order $M^{8/3}$. This is possible if, when the black hole becomes sub-Planckian, the information is stored not in a local Planck-sized region around the origin, but  rather non-locally in a region of radius  $M^{8/3}$: (using $S\sim (ER)^{3/4}$ and $E\sim 1$ this accommodates an entropy $S\sim M^2$). Of course, according to semiclassical gravity macroscopic locality would seem, by the standard argument,  to have to be violated in order for the information to get to such a large region. This can be called a ``non-local remnant". 

\noindent {\bf(5) Maximal information return:}  Late time radiation is correlated with early time radiation, and $S_{ent}$ is nearly zero at the (retarded) time when the black hole becomes Planckian. The profile of $S_{ent}$ is the same as it would be for any blackbody burning down to its ground state, as detailed in \page. This of course also seems to violate macroscopic locality. 
\subsec{Some hints}
\noindent{\bf(i)}
An important hint here comes from thinking about what happens in string theory. Formation/evaporation of a  small (relative to the AdS$_5$ radius) black hole in AdS$_5$ can be described as a process in the dual Yang-Mills gauge theory living on the $S^3$ boundary \refs{\jm,\lt}. We know that the gauge theory is unitary and entropy is well-defined, so this rules out (1) and (2) above.  We can also rule out (3). Since the long-lived remnant lifetime is parametrically longer than the 
evaporation time, the existence of such remnants implies that the gauge theory must contain of order $e^{S_{BH}}=e^{4\pi M^2}$ states with masses less than the 
remnant mass. Since the remnant mass is of order the Planck mass and  $M$ can be arbitrarily large, this says there are infinitely many states below any fixed  energy. But we know this is not the case for gauge theory on a sphere so (3) is ruled out. 
(4) on the other hand cannot be so easily ruled out because the non-local remnant lifetime is shorter than the evaporation time and our understanding of the relation of gauge theory and gravity states is imprecise.  Indeed we do not even know what states correspond to a small black hole, so we certainly can not  analyze the evaporation profile.  However in principle these questions seem answerable.

\noindent{\bf(ii)}
More hints come from thinking about the behavior of field theory two point functions  at infinity in the presence of an eternal black hole \jmc.  Semiclassically, these correlators have a thermal exponential decay at large time separations. However such behavior is incompatible with unitarity  if the black hole is in a pure state: there are always exponentially  rare spikes at late times corresponding to Poincare recurrences.  To compute such exponentially rare processes we must include nonperturbative effects. It would be extremely interesting to understand in some exact theory like string theory how one sees such spikes in the nonperturbative semiclassical expansion.

The problem of computing spikes in correlators seems simpler and more well-posed than the information release problem. An understanding of the former would undoubtedly shed significant  light on and give important hints about the latter. However they are  not exactly the same problem. For one thing, the information release problem already appears in perturbation theory, while the spikes are a nonperturbative phenomenon.  Interesting recent work in this direction appears in \spike. 

\newsec{Conclusion}

We have our work cut out for us. 

\bigskip
 \centerline{\bf ACKNOWLEDGEMENTS}

This work was supported in part by DOE grant DE-FG02-96ER40559.  I am indebted to many colleagues for many wonderful conversations on these topics over the years. 

\listrefs

\end
\bye